\documentclass{iopart}

\usepackage{graphicx}
\begin{document}

\title{Glass transition theory based on stress relaxation}
\author{Kostya Trachenko}
\address{Department of Earth Sciences, University of Cambridge,
Downing Street, Cambridge, CB2~3EQ, UK}

\begin{abstract}
We propose that an onset of glass transition can be defined as the
point at which a supercooled liquid acquires the stress relaxation
mechanism of a solid glass. We translate this condition into the
rate equation for local relaxation events. This equation
simultaneously gives two main signatures of glass transition,
stretched-exponential relaxation and the Vogel-Fulcher law. The
proposed theory quantifies system fragility in terms of the number
of retarded local relaxation events and reproduces its correlation
with the non-exponentiality of relaxation and bonding type.
\end{abstract}


If a liquid is cooled down fast enough, it forms glass. At the onset
of glass transformation range, a liquid qualitatively changes its
properties, and the two main features that distinguish it from a
high-temperature liquid are the stretched-exponential relaxation
(SER) and the Vogel-Fulcher (VF) law. When a perturbation, in the
form of stress or external field, is applied to a liquid near glass
transition, a relaxing quantity $q(t)$ decays following a universal
SER:

\begin{equation}
q(t)\propto\exp(-(t/\tau)^\beta)
\label{eq1}
\end{equation}
\noindent where $0<\beta<1$. This behaviour is seen in many systems,
and is considered a signature of the ``glassy'' relaxation
\cite{kohl,jcp,ngai,bohmer,angell}. Another universal feature of
this regime is that viscosity, or relaxation time $\tau$, follows
non-Arrhenius dependence, which in most cases is well approximated
by the VF law:

\begin{equation}
\tau\propto\exp(A/(T-T_0))
\label{eq2}
\end{equation}
\noindent where $A$ and $T_0$ are constants \cite{angell}.

A substantial amount of research in the area has revolved around the
origin of these two anomalous, yet universal, relaxation laws. A
successful theory of the glass transition, as widely perceived,
should provide a common justification for Eq. (1) and (2)
\cite{ngai,angell10}. Recently, the need for a theory of the glass
transition to give Eq. (1) and (2) {\it simultaneously} has been
reiterated on the basis of the close relationship between $\beta$
and $\tau$: it has been found that $\beta$ is invariant to different
combinations of pressure and temperature that hold $\tau$ constant
\cite{ngainew}. It has therefore been suggested that this
correlation should constrain any theory of the glass transition, in
that if a given formalism gives Eq. (2), it should also be able to
give Eq. (1) \cite{ngainew}.

Several decades ago, Goldstein proposed \cite{gold} that at glass
transition, flow becomes dominated by potential barriers which are
high compared to thermal energies, whereas at high temperature, the
opposite is true, barriers are much smaller than thermal energies.
Hence it can be argued that while in a liquid at high temperature,
local stress is relaxed on the timescale of microscopic trajectory
reversal times, in the supercooled regime, local regions can support
a finite stress (i. e. maintain local structure unchanged) on
timescales that are considerably larger. This has opened the
possibility to discuss the stress relaxation mechanism in a liquid
approaching glass transition. However practical realizations of this
approach, in particular the relationship between the stress
relaxation mechanism and Eq. (1) and (2) have remained elusive.

Following this approach, we consider that as temperature decreases,
liquid acquires a ``solid-like'' ability to support local stresses
on timescales that considerably exceed trajectory reversal times.
Our main proposal is that a liquid near glass transition also starts
to {\it redistribute} local stresses in a solid-like manner. Hence
the onset of the glass transition can be robustly defined as the
point at which a liquid and a solid glass under stress begin to
redistribute local stresses in the same way. In other words, we
propose that {\it the onset of glass transition is the point at
which the liquid acquires the stress relaxation mechanism of solid
glass}. We show that this condition is {\it sufficient} to recover
both anomalous relaxation laws, Eq. (1) and (2). We also show that
the proposed theory gives a simple definition of a system's
fragility in terms of the number of local relaxation events induced
by external perturbation, recovers fragility plots and predicts
correlations of fragility with $\beta$ and the nature of the
chemical bond.

How do local stresses redistribute in glass under pressure? In the
same paper \cite{gold}, Goldstein considered this question: he
argued that because a local region supports less stress after the
relaxation event than before, all other local regions support more
of the external stress after the event than before. Generally,
increasing stress on other local regions makes their relaxation more
difficult. The increased stress they need to support is aligned
along the direction of external pressure. On the other hand, local
relaxation paths for local events with the smallest barriers are
generally oriented at random relative to the external pressure,
because they are defined by the symmetry of local ordering
\cite{trac1} (in Goldstein's terminology, local reaction paths are
``non-concordant'' to the external stress \cite{gold}). As a result,
activation barriers increase for later local relaxation events.

In what follows, we consider that relaxation proceeds by local
relaxation events (LRE). In the literature, these jump or flow
events have been given different names (for review, see Ref.
\cite{dyre2}); in this discussion we borrow the term LRE from our
previous studies of relaxation in glasses
\cite{trac1,trac2,trac21,trac3}. In glass under high enough
pressure, a LRE involves several localized atomic jumps which
include breaking old bonds, forming new ones and the subsequent
relaxation of the local structure \cite{trac1}. An animation of a
LRE in SiO$_2$ glass is available in the electronic form of Ref.
\cite{trac1}. Each LRE carries a microscopic change of a macroscopic
relaxing quantity, e.g., volume. By considering the dynamics of LRE
and their coupling to structural rigidity of glass, it has been
possible to explain several interesting aspects of glass relaxation,
including the origin of slow relaxation \cite{trac2,trac21} and the
origin of temperature-induced densification in the pressure window,
centered at the rigidity percolation point \cite{trac3}.

We introduce LRE as local relaxation ``quanta'' which a liquid uses
to adjust to external perturbations. Each LRE carries a microscopic
change of a liquid's relaxing quantity (i. e. volume, external
stress etc). In a high-temperature liquid, a LRE is an atomic jump
from the surrounding ``cage'', followed by local relaxation. As
temperature decreases, atomic rearrangements become more
cooperative, due to the need to cross higher activation barriers. In
this regime, each LRE is associated with the transition over the
activation barrier in the Goldstein picture of activated flow
\cite{gold}. In the supercooled regime and below, LRE, induced in
different parts of a system, have different relaxation times, i.e.
they are dynamically heterogeneous \cite{hetero}, as discussed below
in more detail.

We now derive the rate equation for LRE in a liquid at the onset of
glass transition. For this, we use our main proposal that at glass
transition, liquid acquires the stress relaxation mechanism of solid
glass. So first we find how to express the stress relaxation
mechanism in glass in mathematical terms. In particular, we find how
activation barriers for LRE change as a result of redistribution of
local stresses. Let $N$ be the total number of relaxing units in the
structure. Under external (hydrostatic or shear) stress $P$, each
unit supports stress $p_0$ such that $P=p_0 N$. Since, as discussed
above, after relaxation, a local unit supports stress $p_1<p_0$, the
stress on other local regions is $p_2=(P-p_1 n_c)/(N-n_c)$, where
$n_c$ is the current (instant) number of LRE induced by external
perturbation. If $n=n_c/N\ll 1$, $p_2=p_0+(p_0-p_1)n$. It has been
argued that the main contributor to the activation barrier $V$ is
elastic energy \cite{dyre2}. Hence the increase of $V$ is
proportional to the increase of work needed to overcome the barrier
created by elastic force due to additional stress $\Delta
p=p_2-p_0=(p_0-p_1)n$. So $V\propto\Delta p\propto n$ for small $n$:

\begin{equation}
V(n)=V_0+V_1 n \label{eq3}
\end{equation}
\noindent where $V_0$ is the energy barrier in an unperturbed
system, and $V_1$ is defined such that $V(n_r$) is the maximal
energy barrier, where $n_r$ is the total number of LRE caused by an
external perturbation, $n(t)\rightarrow n_r$ as
$t\rightarrow\infty$.

Note that Eq. (\ref{eq3}) can not be applied to a liquid above glass
transition, because at high temperature externally-induced stresses
are quickly removed by thermal fluctuations, and redistribution of
stresses between different local regions does not take place. More
precisely, at high temperature, stress relaxation mechanism, as
described by Eq. (\ref{eq3}), only exists on short microscopic
timescales. On experimental timescales, over which Eqs.(\ref{eq1})
and (\ref{eq2}) are measured, $V$ is independent of $n$ at high
temperature.

We now apply Eq. (\ref{eq3}) to the liquid approaching glass
transition. First, the rate of LRE, $\frac{{\rm d}n}{{\rm d}t}$,
depends on the event probability, $\exp(-V/kT)$. According to our
main proposal, the onset of the glass transition is defined as the
point at which a liquid acquires the stress relaxation mechanism of
a solid glass; hence $V$ is given by Eq. (\ref{eq3}). Second,
because an external perturbation induces a finite number of
relaxation events, $n_r$, the rate of LRE should also have a
saturation term to reflect the depletion of LRE. The most natural
choice for the saturation term is linear $-\alpha n$ dependence,
which reflects the fact that relaxed events are removed from further
dynamics. This is analogous to, for example, the process of nuclear
decay, in which the decay rate decreases linearly with the number of
decayed nuclei, ${\rm d}n/{\rm d}t\propto -n$. Hence, using Eq.
(\ref{eq3}) and assuming that in a liquid $V_0 \ll kT$, we write

\begin{eqnarray}
\frac{{\rm d}n}{{\rm d}t}=\exp(-Cn)-\alpha n
\nonumber
\label{eq30}
\end{eqnarray}

\noindent where $C=V_1/kT$ and $t$ is re-scaled as $t\rightarrow
t/t_0$, where $t_0$ is the characteristic relaxation time. $\alpha$
is defined from the condition that $\frac{{\rm d}n_r}{{\rm d}t}=0$
when $n=n_r$, giving $\alpha=\exp(-Cn_r)/n_r$:

\begin{equation}
\frac{{\rm d}n}{{\rm d}t}=\exp(-Cn)-\frac{n}{n_r}\exp(-Cn_r)
\label{eq31}
\end{equation}
\noindent

Before solving Eq. (\ref{eq31}), we note that the only assumption in
its derivation is that the increase of the activation barrier is
linear, Eq. (\ref{eq3}). We argued that this is justified for $n\ll
1$, and here we note that Eq. (\ref{eq3}) is consistent with
experiments in common glasses. If the saturation effects are small
and the second term in the right part of Eq. (\ref{eq31}) is
ignored, $n\propto\ln(t+t_0)$. Thus linear expansion (\ref{eq3})
gives logarithmic dependence of $n$, which is consistent with the
logarithmic relaxation of macroscopic properties (e.g., volume) of
SiO$_2$ and GeO$_2$ glasses under pressure \cite{brazhkin}. In
addition to the qualitative agreement, Eq. (\ref{eq3}) also gives a
quantitative agreement with the experiment, as is found by the
calculation of the slope of logarithmic relaxation of volume under
pressure \cite{trac2}.

Eq. (\ref{eq31}) has two parameters, $C$ and $n_r$. When $Cn_r\ll 1$
(and hence $Cn\ll 1$ since $n<n_r$), the right part of Eq.
(\ref{eq31}) becomes $1-n/n_r$, leading to the usual Debye- type
relaxation. This takes place when either the temperature $T\propto
1/C$ is high, or when $n_r$ is small. At $Cn_r\approx 1$, one
expects the onset of non-exponential relaxation. This sets the scale
for the non-exponentiality temperature $T_n$:

\begin{equation}
kT_n \approx V_1 n_r \label{eq4}
\label{eq4}
\end{equation}

Below we show that in the $T\approx T_n$ regime, Eq.(4) gives SER
and the VF law, Eqs. (1) and (2). Note that $T_n$ is higher than the
glass transition temperature $T_g$. $T_g$ is often defined from the
condition of viscosity reaching some large value, corresponding to
the relaxation time exceeding the time of experiment, whereas $T_n$
defines the preceding temperature regime at which relaxation becomes
non-exponential and non-Arrhenius.

First, we solve Eq. (\ref{eq31}) for a wide range of parameters ($C,
n_r$) that satisfy condition $Cn_r\geq 1$, i.e. define the
non-exponential regime of solution of Eq. (\ref{eq31}). Remembering
that $n(t)\rightarrow n_r$ as $t\rightarrow\infty$, we fit the
solution to $n=n_r(1-\exp(-t/\tau)^\beta)$. We note that this form
of SER and Eq. (\ref{eq31}) contain two parameters each, which
suggests that if a good fit exists, it is not accidental, but
probably reflects the involved physics. Figure 1 shows that fits of
the solution to SER are very good. We find that this is the case in
the wide range of parameters ($C, n_r$), except when $Cn_r\gg$1.

An important observation from Figure 1 (see the legend) is that
$\beta$ decreases as $Cn_r$ increases. Hence non-exponentiality can
increase as a result of either increase of $n_r$, or decrease of
temperature $T\propto 1/C$. This is consistent with many experiments
in which $\beta$ decreases with $T$ \cite{jcp,ngai,bohmer}. We will
return to this point below.

\begin{figure}
\begin{center}
\rotatebox{-90}{\scalebox{0.47}{\includegraphics{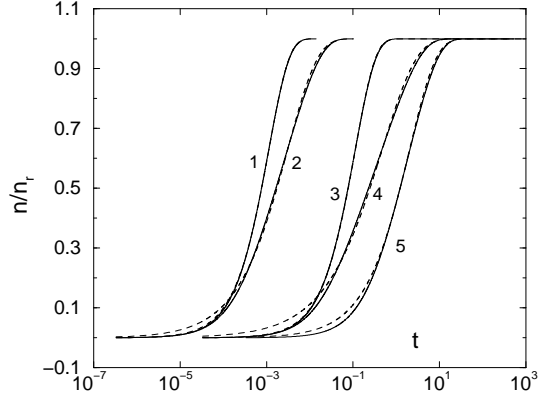}}}
\end{center}
\caption{Solid lines are the solutions of Eq. (4) for several pairs
of parameters ($n_r,C$): 1 - (0.001,1000), 2- (0.001,4000), 3-
(0.1,10). 4 - (0.1,50), 5 - (1,3). Dashed lines are the least-square
fits to SER, giving the following parameters of ($\beta$, $\tau$):
1- (0.93,0.0011), 2 - (0.63,0.0031), 3- (0.94,0.114), 4 -
(0.55,0.478), 5 - (0.71,2.05), respectively. For each value of
$n_r$, the solution of Eq.(4) for $n$ has been divided by $n_r$ so
that 0$<n/n_r<$1.} \label{fig1}
\end{figure}

Second, for several different $n_r$, we solve Eq.(4) as a function
of $C$, and fit the solution to the SER form above to obtain $\tau$.
We plot the solution as a function of 1/($n_r C$)$=T/T_n$, where
$T_n$ is defined from Eq. (\ref{eq4}). We find that the dependence
of relaxation time $\tau$ on $T/T_n$ collapses on the curve
ln($\tau/n_r$)=$f$($T/T_n$). We also find that $f(x)$ can not be
represented by the Arrhenius-type dependence $\propto 1/x$, however
a good fit is obtained if

\begin{equation}
\ln\frac{\tau}{n_r}=\frac{a_1}{1/(Cn_r)-a_2}=\frac{a_1}{T/T_n-a_2}
\label{eq5}
\end{equation}
\noindent where $a_1$ and $a_2$ are constants. This is the form of
the VF law, Eq. (\ref{eq2}). We find that a good fit to Eq.
(\ref{eq5}) exists in both $T>T_n$ and $T<T_n$ regimes, with
ln($\tau/n_r$) spanning over 15 decades (see Figure 2).

\begin{figure}
\begin{center}
\rotatebox{-90}{\scalebox{0.4}{\includegraphics{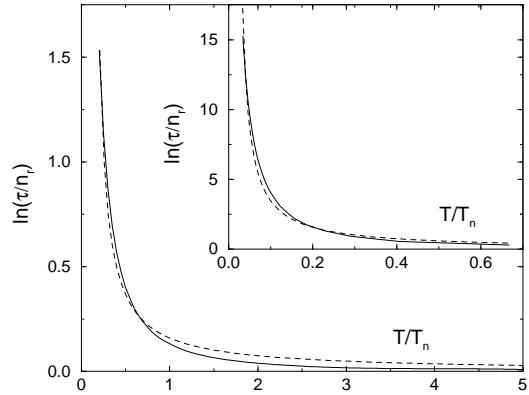}}}
\end{center}
\caption{The solid line is the solution of Eq. (4), fitted to SER to
obtain $\tau$. The dashed line is the fit to Eq. (6), with
$a_1$=0.141 and $a_2$=0.116. The insert shows the fit in the $T<T_n$
regime; $a_1$=0.288 and $a_2$=0.017.}
\label{fig2}
\end{figure}

The transition from Eq. (\ref{eq5}) to Arrhenius dependence directly
follows from the condition that solution of Eq. (\ref{eq31}) is
exponential, $Cn_r \ll 1$. Since $a_2<1$ (see the legend to Figure
2), this condition means $1/(Cn_r)\gg a_2$, and dependence
(\ref{eq5}) becomes Arrhenius. If, on the other hand, $Cn_r \ll 1$
does not hold, relaxation is VF-type.

We now suggest that the proposed theory clarifies the origin of
system ``fragility'', its correlation with the non-exponentiality of
relaxation \cite{angell10,bohmer} and the nature of the chemical
bond.

First, $n_r$, a parameter in Eq. (\ref{eq31}), quantifies the
overall atomic motion in a system due to external perturbation that
comes in addition to thermally-induced motion. At a given
temperature, $n_r$ depends on the magnitude of external perturbation
and, more importantly, on the system's ability to resist structural
changes at the microscopic level. This ability has been termed a
system ``fragility'', and constitutes the basis of fragility plots,
which are essentially plots of Eq. (\ref{eq5}) with a varying
parameter that measures the deviation from Arrhenius dependence; the
larger this deviation, the larger the fragility \cite{angell}.

Qualitatively, a ``strong'' system has a built-in resistance to
temperature-induced structural changes, whereas the structure of a
``fragile'' system is easy to disrupt \cite{angell}. In our picture,
this means that a strong system responds to external perturbation
with little retardation and smaller $n_r$, whereas relaxation in a
more fragile system involves a larger number of LRE, required to
come to equilibrium with new external conditions. In other words, a
more fragile system is more retarded in terms of larger $n_r$.

Hence our picture offers the quantification of fragility in terms of
$n_r$, and we can immediately recover fragility plots in this
approach. As discussed above, small $n_r$ corresponds to a stronger
system. At a given temperature, small $n_r$ results in $1/(Cn_r)\gg
a_2$, and Eq. (\ref{eq5}) becomes Arrhenius. As $n_r$ increases,
corresponding to a more fragile system in our picture, $1/(Cn_r)\gg
a_2$ does not hold, and relaxation becomes progressively
non-Arrhenius (see Eq. (\ref{eq5})). Another way of discussing this
effect is to note that $n_r$ is proportional to $T_n$ (see Eq.
\ref{eq4}); hence larger $n_r$ in a fragile system corresponds to
higher $T_n$. Physically, this means that the increase of the degree
of system's retardation, quantified by $n_r$, requires a higher
temperature to remove this retardation by a more efficient
equilibration and make the system relax exponentially.

Second, we find that the proposed picture reproduces the
relationship between fragility and non-exponentiality. Experimental
data of more than 70 systems \cite{bohmer} show that $\beta$
decreases linearly with fragility. In our picture, fragility is
defined by $n_r$, and in Figure 3, we plot $\beta$ as a function of
$n_r$ for different values of $C$. It is indeed seen that $\beta$
decreases with $n_r$, reproducing the experimental correlation well.
Note that at a given $n_r$, higher $T\propto 1/C$ results in the
increase of $\beta$ (see Figure 3), in agreement with experimental
observations \cite{jcp,ngai,bohmer}.

\begin{figure}
\begin{center}
\rotatebox{-90}{\scalebox{0.4}{\includegraphics{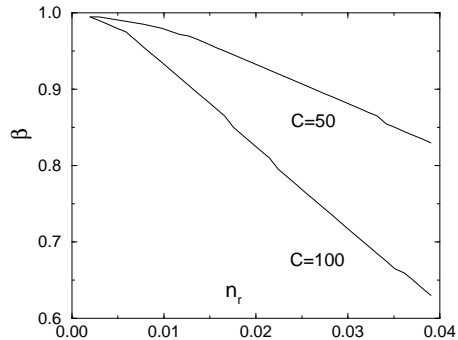}}}
\end{center}
\caption{Decrease of $\beta$ with $n_r$ (fragility), at various
values of $C$. At constant $n_r$, $\beta$ increases with $T\propto
1/C$. } \label{fig3}
\end{figure}

Finally, one can discuss how the proposed picture relates a system's
fragility to its microscopic parameters. Other conditions being
equal, one expects that an external perturbation induces generally
larger $n_r$ in a system with ionic bonding as compared with a
system with covalent bonding. In the covalent case, an atomic pair
lowers its energy through sharing electrons between two atoms,
resulting in a binding energy as high as several eV, and a LRE
necessarily involves breaking these stable electronic configurations
(breaking ``covalent bonds'') with associated high energy cost. In
the ionic case, atomic rearrangements can proceed without a change
in the electronic state of the atoms. As a result, activation
barriers generally increase with covalency of bonding. Since, as
discussed above, fragility increases with $n_r$, one readily
predicts that covalent systems should be generally stronger and
ionic systems should be more fragile, in good agreement with
experimental results \cite{angell}. Other factors, in addition to
the nature of the chemical bond, may also affect $n_r$, including
for example, the ratio of ionic radii.

Before concluding, we make three remarks. First, it is important to
note that Eq. ({\ref{eq31}) yields SER and the VF law {\it
simultaneously}, suggesting that {\it LRE dynamics are behind both
anomalous ``glassy'' relaxation laws} that kick in at the onset of
the glass transition. This clarifies an open question of why the
relaxation function is non-exponential at temperatures at which the
relaxation time is non-Arrhenius \cite{angell10}: in our theory,
larger $n_r$ increases the non-exponentiality of relaxation (see
Figures 1 and 3) and, at the same time, increases departure from the
Arrhenius relaxation as follows from Eq. (\ref{eq5}).

A second related point is that the relationship between SER and the
VF law has recently been reiterated: it has been discovered that
$\beta$ is invariant to different combinations of pressure and
temperature that hold $\tau$ constant \cite{ngainew}. It has
therefore been suggested that this correlation should constrain any
theory of the glass transition \cite{ngainew}. In our theory,
temperature and pressure define parameters $C$ and $n_r$ in Eq.
(\ref{eq31}). $C$ and $n_r$, in turn, unambiguously define $\beta$
and $\tau$. Due to the monotonous character of the solution of Eq.
(\ref{eq31}), we find that only one value of $\beta$ corresponds to
a given $\tau$. In other words, we find that $\beta$ is invariant to
different combinations of pressure and temperature that hold $\tau$
constant, satisfying the experimental result \cite{ngainew}.

Finally, we note that in order to derive Eq. (1) and (2), we
considered a system under external stress, which allowed us to
discuss the stress relaxation mechanism near glass transition, Eq.
(\ref{eq3}-\ref{eq31}). At the same time, Eq. (1) and (2) are
observed in supercooled liquids in the absence of pressure as well,
from decay of correlation functions. This behaviour directly follows
from the considered situation of the system under stress, by
applying the fluctuation-dissipation theorem.

In summary, we proposed a new simple way of defining the onset of
glass transition: a liquid enters glass transformation range when it
begins to redistribute local stresses in the same manner as solid
glass. We showed how this condition simultaneously gives two main
signatures of ``glassy'' relaxation, the stretched-exponential
relaxation and the Vogel-Fulcher law. Consistent with recent
experiments, we found that in the proposed theory, $\beta$ is
invariant to different combinations of pressure and temperature that
hold $\tau$ constant. We have discussed that the proposed theory
offers the definition of system's fragility in terms of the number
of local relaxation events, and recovers experimental correlations
of fragility with non-exponentiality of relaxation and the nature of
the chemical bond.

\ack I am grateful to Prof. J. C. Phillips, R. B\H{o}hmer, D. L.
Stein, M. T. Dove and V. V. Brazhkin for discussions, and to EPSRC
for support.


\begin{thebibliography}{99}

\bibitem{kohl} Kohlrausch R 1854 {\it Pogg. Ann. Phys. Chem.} {\bf 91} 56/179

\bibitem{jcp} Phillips J C 1996 {\it Rep. Prog. Phys.} {\bf 59} 1133

\bibitem{ngai} Ngai K L 2000 {\it J. Non-Cryst. Sol.} {\bf 275} 7

\bibitem{bohmer} B\H{o}hmer R et al 1993 {\it J. Chem. Phys.} {\bf 99} 4201

\bibitem{angell} Angell C A 1995 {\it Science} {\bf 267} 1924

\bibitem{angell10} Angell C A 2000 {\it J. Phys.: Cond. Matt.} {\bf 12} 6463

\bibitem{ngainew} Ngai K L et al 2005 {\it J. Phys. Chem. B} {\bf 109} 17356

\bibitem{gold} Goldstein M 1969 {\it J. Chem. Phys.} {\bf 51} 3728

\bibitem{trac1} Trachenko K and Dove M T 2002 {\it J. Phys.: Cond. Matt.} {\bf 14} 7449

\bibitem{dyre2} Dyre J C 1998 {\it J. Non-Cryst. Sol.} {\bf 235-237} 142 and
references therein.

\bibitem{trac2} Trachenko K, Dove M T, Brazhkin V V and Phillips J C 2003 {\it J. Phys.: Cond. Matt}
{\bf 15} L743

\bibitem{trac21} Trachenko K and Dove M T 2004 {\it Phys. Rev. B} {\bf 70} 132202

\bibitem{trac3} Trachenko K et al 2004 {\it Phys. Rev. Lett.} {\bf 93} 135502; Trachenko K and Dove M T 2003
{\it Phys. Rev. B} {\bf 67} 212203

\bibitem{hetero} B\H{o}hmer R et al 1998 {\it J. Non. Cryst. Sol.} {\bf 235-237} 1

\bibitem{brazhkin} Tsiok O B et al 1998 {\it Phys. Rev. Lett.} {\bf 80} 999

\end{thebibliography}
\end{document}